# Randomization for Security in Half-Duplex Two-Way Gaussian Channels

Aly El Gamal, Moustafa Youssef
Wireless Intelligent Networks
Center (WINC)
Nile University, Cairo, Egypt
ali.melgamal@nileu.edu.eg, mayoussef@nileuniversity.edu.eg

Hesham El Gamal
Department of Electrical
and Computer Engineering
Ohio State University, Columbus, USA
helgamal@ece.osu.edu

[1] *Abstract*—This paper develops a new physical layer framework for secure two-way wireless communication in the presence of a passive eavesdropper, i.e., Eve. Our approach achieves perfect information theoretic secrecy via a novel *randomized scheduling and power allocation* scheme. The key idea is to allow Alice and Bob to send symbols at random time instants. While Alice will be able to determine the symbols transmitted by Bob, Eve will suffer from ambiguity regarding the source of any particular symbol. This *desirable* ambiguity is enhanced, in our approach, by *randomizing* the transmit power level. Our theoretical analysis, in a 2-D geometry, reveals the ability of the proposed approach to achieve relatively high secure data rates under mild conditions on the spatial location of Eve. These theoretical claims are then validated by experimental results using IEEE 802.15.4-enabled sensor boards in different configurations, motivated by the spatial characteristics of Wireless Body Area Networks (WBAN).

## I. INTRODUCTION

Recently, there has been a growing interest in the area of physical layer security for wireless networking applications (e.g., [1], [2], [3]). The underlying idea is to exploit the characteristics of the wireless medium to develop communication protocols with provable information theoretic security guarantees. The notion of information theoretic security can be traced back to the pioneering work of Shannon [4] which considered the basic model of a one-way point-to-point communication link where both the sender and the destination possess a common secret key (used to encrypt and decrypt the message). This seminal work introduced the perfect secrecy condition $I(M; Z) = 0$ implying that the signal $Z$ received by the eavesdropper does not provide any additional information about the source message $M$ (i.e., zero mutual information between $M$ and $Z$). Under the assumption that both the eavesdropper and legitimate destination receive the transmitted message through **a noiseless channel**, Shannon proved that the achievability of perfect secrecy requires the entropy of the shared private key $K$ to be at least equal to the entropy of the message itself (i.e., $H(K) \geq H(M)$). The challenging task of distributing/updating secret keys in wireless networks motivates the *key-less* security approach proposed in the sequel.

[1]This research is supported in part by a grant from the Egyptian National Telecommunication Regularity Authority (NTRA) and in part by a grant from British Petroleum.

The proposed scheme builds on the work of Lai *et al.* [5] which developed an approach for using randomized feedback to perform encryption *over modulo additive channels* (e.g., binary channels) without sharing a secret key *a priori*. The basic idea is to use the randomized feedback as a jamming signal for the eavesdropper who, due to the modulo-additive nature of the channel, can not differentiate between the corrupted and non-jammed symbols. Interestingly, this scheme was shown to achieve remarkable secrecy advantage even under the half-duplex constraint whereby the transmission of one jamming symbol results in an erasure at the legitimate receiver. Here, we extend this scheme to the wireless two-way Gaussian channel. The real-valued nature of the channel gives the eavesdropper a certain ability to identify the corrupted symbols, perhaps via a symbol power detector. We overcome this problem via a novel *randomized scheduling and power allocation scheme*. More precisely, in a two-way communication session, the legitimate nodes pick the transmission times randomly without any prior agreement. This results in a certain loss of throughput, due to the half-duplex assumption, when both nodes are transmitting (or not transmitting) simultaneously, but creates a **significant** advantage over the eavesdropper who will find it rather difficult to associate the different symbols with their source node. To further increase the ambiguity at the eavesdropper, the power level used in each symbol is chosen randomly according to a predetermined distribution (known to all nodes). The overall effect is to guarantee provable secrecy of the two transmitted messages at the expense of a minimal loss in throughput. We argue in the sequel that Secure Wireless Body Area Network (SW-BAN) is an ideal candidate application for the proposed approach. The reason is that, in this scenario, the distance between the eavesdropper and any of the two legitimate nodes is expected to be much larger than the inter-node distance; a property that results in maximal ambiguity at the eavesdropper (as shown in the sequel).

The rest of the paper is organized as follows. Section II details our system model and notations. In Section III, the theoretical foundation of the proposed randomized scheduling approach is developed. We report experimental results that validate our theoretical claims in Section IV-B. Finally, Section V offers few concluding remarks.

## II. SYSTEM MODEL

We consider the basic three-terminal setup where two legitimate nodes, i.e., Alice and Bob, wish to interchange messages in the presence of a passive eavesdropper, Eve. Our focus will be devoted to the symmetric case where the two messages have the same rate. Alice and Bob are equipped with a single half-duplex antenna implying that each node can either transmit or receive (but not both) on the same degree of freedom. In our analysis, we employ a $2$-$D$ Geometric model where, without any loss of generality, Alice and Bob are assumed to be located on the $x$-axis at opposite ends of the origin. Motivated by the Body Area Network (BAN) application, Eve is assumed to be located **outside** a circle centered around the origin of radius $r_E$ at an angle $\theta$ of the $x$-axis. This **key** assumption faithfully models the spatial separation, between the legitimate nodes and eavesdropper(s), which characterizes BANs. The performance of the proposed secure randomized scheduling communication scheme will be obtained as a function of $r_E$ and the distance between Alice and Bob, i.e., $d_{AB}$. In our discrete time model, the signals received by the three nodes in the $i^{th}$ symbol interval are given by

$$Y_E(i) = G_E(d_{AE}^{-\alpha/2} X_A(i) e^{-jkd_{AE}} + \quad (1)$$
$$\qquad d_{BE}^{-\alpha/2} X_B(i) e^{-jkd_{BE}})$$
$$\qquad + N_E(i)$$

$$Y_A(i) = (1 - \mathcal{I}(X_A(i)))$$
$$\qquad (G_A(d_{AB}^{-\alpha/2} X_B(i) e^{-jkd_{AB}}) + N_A(i))$$

$$Y_B(i) = (1 - \mathcal{I}(X_B(i)))$$
$$\qquad (G_B(d_{AB}^{-\alpha/2} X_A(i) e^{-jkd_{AB}} + N_B(i)),$$

where $k$ is the wave number, $N_A(i)$, $N_B(i)$, and $N_E(i)$ are the unit-variance zero-mean additive white Gaussian noise samples at Alice, Bob, and Eve respectively. Furthermore, $G_A$, $G_B$ and $G_E$ are propagation constants which depend on the receive antenna gains, and $\alpha$ is the path loss exponent which will be taken to 2 as in the free space propagation scenario (one can easily extend our results for other scenarios with different path loss exponents.). For simplicity, we restrict ourselves to binary encoding implying that $X_A(i) \in \left\{\sqrt{\rho(i)}, 0, -\sqrt{\rho(i)}\right\}$, where $\rho(i)$ is the instantaneous signal to noise ratio at $d_{AB} = 1$ in the $i^{th}$ symbol interval if Alice decides to transmit whereas $X_A(i) = 0$ if Alice decides not to transmit (the same applies to $X_B(i)$). $\rho(i)$ is selected randomly in the range $[\rho_{min}, \rho_{max}]$ according to a distribution that is known *a priori* to all nodes. $\mathcal{I}(X_A(i))$ is the indicator function, i.e., $\mathcal{I}(X_A(i)) = 1$ if $X_A(i) \neq 0$ and zero otherwise. Therefore, $Y_A(i) = 0$ if $X_A(i) \neq 0$ as dictated by the half-duplex assumption.

In order to invoke information theoretic arguments, we assume an asymptotically long frame length $T \to \infty$. Moreover, in order to ensure the robustness of our results, we assume that Eve employs a large enough receive antenna, i.e., $G_E >> 1$, such that the additive noise effect in $Y_E$ can be ignored. We further assume a hard decision decoder at both the legitimate receiver and the eavesdropper, and a *memoryless* strategy $\mathcal{C}$ is assumed to be used by the classifier employed at Eve to identify the origin of each received symbol. i.e. the decision is based only on the power level of the observed symbol in the current time interval. Finally, we use the following notations: $[a]^+ = \max(a, 0)$, $\phi(x) = \int_{-\infty}^{x} \frac{1}{\sqrt{2\pi}} e^{\frac{-t^2}{2}} dt$.

## III. SECURE TWO-WAY COMMUNICATION

### A. One-Way Communication with Feedback

Inspired by the earlier work of Lai *et al* [5], we first consider a **Time Division Multiplexing** scheme whereby only a single message transfer takes place in any time frame, and the legitimate receiver *jams* the channel with random-content feedback symbols at random time intervals. The receiver will transmit a feedback symbol at any time interval with probability $\beta$. The randomized schedule of feedback will result in erroneous outputs at the eavesdropper due to its inability to identify the symbols corrupted by the random feedback signal. As argued in [5], this scheme is capable of completely impairing Eve in modulo-additive channels. In our *real-valued* channel, however, a simple energy classifier based on the average received signal power [6] can be used by Eve to differentiate between corrupted and non *jammed* symbols. To overcome this problem, we use predetermined distributions for the transmit power of both the data symbols, $f_1$, and feedback symbols $f_2$. This randomized power allocation strategy is intended to increase the probability of *misclassification* at Eve. Assuming that the classifier task is to *detect* the presence of the feedback signal, we use $P_m$ and $P_f$ to represent the probability of miss-detection and false alarm; respectively. Furthermore, we use $P_{e|m}$ to denote the probability of symbol error given occurrence of the miss-detection event. The following result provides a lower bound on the achievable secrecy rate using this protocol.

*Theorem 1:* Using the proposed one-way protocol with randomized feedback and power allocation, the achievable secrecy rate is lower bounded by:

$$R_s \geq 0.5 \max_{\beta, f_1, f_2} (\min_{\theta, \mathcal{C}}([R_M - R_E]^+)) \quad (2)$$

where:

$$R_M = (1 - \beta)\left(1 - H\left(1 - \phi\left(\sqrt{\frac{\rho_{min}}{d_{AB}^\alpha}}\right)\right)\right)$$

$$R_E = (1 - \beta(1 - P_m) - (1 - \beta)P_f)$$
$$\qquad \left(1 - H\left(\frac{\beta P_m P_{e|m}}{1 - \beta(1 - P_m) - (1 - \beta)P_f}\right)\right)$$

*Proof:* Let $\alpha_M, \alpha_E$ denote the fraction of symbols erased at Bob and Eve, and $P_e^{(M)}, P_e^{(E)}$ denote the probability of erroneously decoding a received symbol given that it was not erased at Bob and Eve, respectively. By applying the appropriate random binning scheme [7], the following secrecy rate is achievable ([8], Theorem 3):

$$R_s = \max_{p(x)}([(I(X;Y) - I(X;Z))]^+)$$

where $X$ denotes the input, $Y$ and $Z$ denote the outputs at Bob and Eve respectively. Considering the transition model for this channel, it is straightforward to see that:

$$H(Y|X) = H(\alpha_M) + (1-\alpha_M)H(P_e^{(M)})$$

Now Let $P_{X=1} = \Pi$ then

$$H(Y) = H(\alpha_M) + (1-\alpha_M)H(\Pi(1-P_e^{(M)}) + (1-\Pi)P_e^{(M)}),$$

and

$$\max_\Pi H(Y) = H(\alpha_M) + (1-\alpha_M),$$

when $\Pi = 0.5$. This results in

$$\max_{p(x)} I(X;Y) = \max_{p(x)}(H(Y) - H(Y|X))$$
$$= (1-\alpha_M)(1-H(P_e^{(M)}))$$

Similarly,

$$\max_{p(x)} I(X;Z) = (1-\alpha_E)(1-H(P_e^{(E)}))$$

Following the half-duplex assumption, all data symbols transmitted during the same time interval of a feedback transmission will be considered as erasures at the legitimate receiver's channel. Therefore, as the frame length $T \to \infty$, $\alpha_M = \beta$. For the rest of the symbols, the probability of symbol error by the hard decision detector will be

$$P_e^{(M)}(i) = 1 - \phi\left(\sqrt{\frac{\rho(i)}{d_{AB}^\alpha}}\right).$$

On the other hand, feedback transmissions will introduce decoding errors at the eavesdropper. Noting that $1-P_m$ of those corrupted symbols will be detected by the energy classifier, we get

$$\alpha_E = \beta(1-P_m) + (1-\beta)P_f$$

$$P_e^{(E)} = \frac{\beta P_m P_{e|m}}{1-\alpha_E}.$$

Combining these results, we obtain

$$\max_{p(x)} I(X;Y) = \max_{p(x)} H(Y) - H(Y|X)$$
$$= (1-\alpha_M)(1-H(P_e^{(M)}))$$
$$= (1-\beta)\left(1 - H\left(\frac{1}{T}\sum_{i=1}^T P_e^{(M)}(i)\right)\right)$$
$$\geq (1-\beta)\left(1 - H\left(1 - \phi\left(\sqrt{\frac{\rho_{min}}{d_{AB}^\alpha}}\right)\right)\right)$$
$$= R_M$$

$$\max_{p(x)} I(X;Z) = \max_{p(x)} H(Z) - H(Z|X)$$
$$= (1-\alpha_E)(1-H(P_e^{(E)}))$$
$$= (1-\beta(1-P_m) - (1-\beta)P_f)$$
$$\left(1 - H\left(\frac{\beta P_m P_{e|m}}{1-\alpha_E}\right)\right)$$
$$= R_E$$

$$R_s = \max_{p(x)}([I(X;Y) - I(X;Z)]^+)$$
$$\geq ([\max_{p(x)} I(X;Y) - \max_{p(x)} I(X;Z)]^+)$$
$$\geq ([R_M - R_E]^+)$$

Finally, we consider a *max-min* strategy whereby the legitimate receiver assumes that the eavesdropper chooses its position around the perimeter of the circle and the energy classifier's mechanism $\mathcal{C}$ to minimize the secrecy rate $R_s$. Accordingly, the legitimate receiver determines the probability of random feedback transmission $\beta$ and both the data and feedback signal power distributions $f_1$, $f_2$ to maximize this worst case value (note that the rate is scaled by $0.5$ to account for the time division between the two nodes):

$$R_{sec} = 0.5 \max_{\beta, f_1, f_2} (\min_{\theta, \mathcal{C}} R_s)$$

∎

### B. Two-Way Communication with Randomized Scheduling

Unlike the scheme described in section III-A where a prior agreement on TDM frames is required, we now propose a *two-way communication* protocol where both legitimate nodes exchange messages via a randomized scheduling protocol. In this scheme, each node will transmit its message in randomly selected time intervals, where a single node's transmitter is active in any given time interval with probability $P_t$, and the transmit power level is randomly selected according to the distribution $f$. Consequently, there are four possibilities for the status of both transmitters in any particular time interval $i$. Due to our noiseless assumption, the eavesdropper's antenna will easily identify *silence* intervals. The challenge, facing the eavesdropper classifier, is to differentiate between the 3 other events. Let $A$ and $B$ represent the activation of Alice's and Bob's transmitters; respectively, $A^c$ and $B^c$ represent their deactivation, and $E1 \to E2$ represents the occurrence of event $E1$ and its classification by Eve as event $E2$. Moreover, we let $P_{e|(A,B) \to (A,B^c)}$ denote the probability of error given that the event $(A,B)$ was mistaken for $(A,B^c)$ by the classifier. The following is our main result in this section

*Theorem 2:* Using the proposed **randomized scheduling and power allocation protocol**, the achievable secrecy rate is lower bounded by:

$$R_{sec} \geq \max_{P_t, f}(\min_{\theta, \mathcal{C}}([R_M - \max(R_{EA}, R_{EB})]^+)) \quad (3)$$

where:

$$R_M = P_t(1-P_t)\left(1 - H\left(1 - \phi\left(\sqrt{\frac{\rho_{min}}{d_{AB}^\alpha}}\right)\right)\right)$$

$$R_{EA} = D_A\left(1 - H\left(\frac{P_e^{(EA)}}{D_A}\right)\right)$$

$$R_{EB} = D_B\left(1 - H\left(\frac{P_e^{(EB)}}{D_B}\right)\right)$$

$$D_A = P_t^2 P_{(A,B)\to(A,B^c)} + P_t(1-P_t)P_{(A^c,B)\to(A,B^c)}$$
$$+ P_t(1-P_t)\left(1 - P_{(A,B^c)\to(A^c,B)} - P_{(A,B^c)\to(A,B)}\right)$$

$$D_B = P_t^2 P_{(A,B)\to(A^c,B)} + P_t(1-P_t)P_{(A,B^c)\to(A^c,B)}$$
$$+ P_t(1-P_t)\left(1 - P_{(A^c,B)\to(A,B^c)} - P_{(A^c,B)\to(A,B)}\right)$$

$$P_e^{(EA)} = P_t^2 P_{(A,B)\to(A,B^c)} P_{e|(A,B)\to(A,B^c)}$$
$$+ 0.5 P_t(1-P_t) P_{(A^c,B)\to(A,B^c)}$$

$$P_e^{(EB)} = P_t^2 P_{(A,B)\to(A^c,B)} P_{e|(A,B)\to(A^c,B)}$$
$$+ 0.5 P_t(1-P_t) P_{(A,B^c)\to(A^c,B)}$$

where $D_A$, $D_B$ represent the portion of symbols classified by Eve as being transmitted by Alice or Bob respectively.

*Proof:* Due to symmetry, we only consider the secrecy rate of Alice's message to Bob. Following in the footsteps of [7], one can argue that:

$$\begin{aligned}R_s &= \max_{p(x)}\left([I(X;Y) - I(X;Z)]^+\right) \\ &\geq \left[\max_{p(x)} I(X;Y) - \max_{p(x)} I(X;Z)\right]^+ \\ &= [(1-\alpha_M)(1-H(P_e^{(M)})) \\ &\quad -(1-\alpha_E)(1-H(P_e^{(E)}))]^+\end{aligned}$$

where $\alpha_M, \alpha_E, P_e^{(M)}$, and $P_e^{(E)}$ are defined as in the proof of Theorem 1. Using half-duplex antennas, each node will be able to decode a symbol transmitted by the other node only when its own transmitter is idle and the other node's transmitter is active. These two conditions are simultaneously satisfied with probability $P_t(1-P_t)$ yielding $\alpha_M = 1 - P_t(1-P_t)$. It is also straightforward to see that

$$P_e^{(M)}(i) = 1 - \phi\left(\sqrt{\frac{\rho(i)}{d_{AB}^\alpha}}\right)$$

The symbols classified by Eve as being transmitted by Alice can belong to one of three categories. The first, which takes place with probability $P_t(1-P_t)\left(1 - P_{(A,B^c)\to(A^c,B)} - P_{(A,B^c)\to(A,B)}\right)$, represents the portion successfully detected and correctly decoded by the eavesdropper. The second corresponds to symbols transmitted by Bob and misclassified as belonging to Alice; with probability $P_t(1-P_t)P_{(A^c,B)\to(A,B^c)}$. Those symbols are independent from the ones transmitted by Alice, and hence, have a probability 0.5 of being different. The third category, with probability $P_t^2 P_{(A,B)\to(A,B^c)}$, corresponds to concurrent transmissions that are not *erased* by Eve's classifier and misclassified as Alice's symbols. The probability of error in these symbols is denoted by $P_{e|(A,B)\to(A,B^c)}$. Combining these terms, we get

$$\alpha_E = 1 - D_A$$

$$P_e^{(E)} = \frac{P_e^{(EA)}}{1 - \alpha_E}$$

$$\begin{aligned}R_s &\geq \max_{p(x)} I(X;Y) - \max_{p(x)} I(X;Z) \\ &= [(1-\alpha_M)(1-H(P_e^{(M)})) - \\ &\quad (1-\alpha_E)\left(1 - H\left(P_e^{(E)}\right)\right)]^+ \\ &\geq [P_t(1-P_t) \\ &\quad \left(1 - H\left(1 - \phi\left(\sqrt{\frac{\rho_{min}}{d_{AB}^\alpha}}\right)\right)\right) \\ &\quad - D_A\left(1 - H\left(P_e^{(E)}\right)\right)]^+\end{aligned}$$

Finally, we follow the same min-max strategy as the proof of Theorem 1. ∎

## IV. NUMERICAL AND EXPERIMENTAL RESULTS

### A. Numerical Results

In the following, we evaluate both schemes under selected assumptions for the transmit signal power distribution and Eve's classifier. We assume the same uniform power distribution for both Alice and Bob, and a threshold based energy classifier with two limits, namely $T_1$ and $T_2$, is used by Eve. Using this classifier, a received symbol is erased if the received signal power falls outside the interval between $T_1$ and $T_2$, otherwise the received symbol is forwarded to the decoder. For the randomized scheduling approach, the decoder is selected according to the following decision rule:

$$\mathcal{D}(R_i) = \begin{cases} A & if \quad A_{min} < R_i < \min(A_{max}, B_{min}) \\ B & if \quad R_i > A_{max} \\ A|B & Otherwise \end{cases}$$

where $A_{min}$ and $A_{max}$ are the minimum and maximum received signal power values in dBm for Alice's transmission, $B_{min}$ and $B_{max}$ are the analogous values for Bob, $R_i$ is the observed RSSI value at Eve for the $i^{th}$ symbol, $A$,$B$ denote classification decisions of the received symbol as being transmitted by Alice or Bob respectively, and $A|B$ means choosing Alice or Bob with equal probabilities. The shown rule is used for the case when $d_{BE} \leq d_{AE}$. Our noiseless assumption implies that Eve will decode the received symbols, corresponding to concurrent transmissions, as the symbols with the higher received signal power. To simplify the presentation, we further assume that Alice and Bob use sufficient error control coding to overcome the additive noise effect. More precisely, Alice and Bob are assumed to use asymptotically optimal forward error control coding and that $\frac{\rho_{min}}{d_{AB}^2}$ is above the minimal power level required to achieve arbitrarily vanishing probability of error at the legitimate receivers.

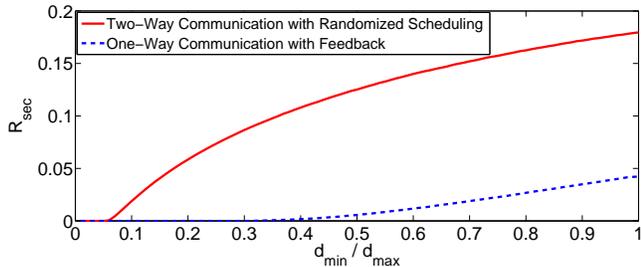

Fig. 1. Maximum achievable secrecy rate for different distance ratios between Eve and each of the two communicating nodes.

Figure 1 reports the lower bounds, on the secrecy rate $R_{sec}$, of Theorems 1 and 2 at different values for the distance ratio $\frac{d_{min}}{d_{max}}$ ($d_{min} = \min(d_{AE}, d_{BE})$, $d_{max} = \max(d_{AE}, d_{BE})$.). The rates plotted for the one-way TDM scheme are obtained assuming that the legitimate transmitter lies at $d_{min}$, i.e. Eve is closer to the transmitter than the receiver. This configuration was found to achieve lower secrecy rates, as a **larger fraction** of the jammed symbols can be correctly decoded at Eve. A few remarks are now in order

1) It is evident that our two-way randomization scheme achieves higher rates than the TDM scheme. The reason is the added ambiguity at Eve resulting from the randomization in the scheduling algorithm.
2) The lower value of $R_{sec}$ achieved for both schemes for smaller values of the ratio $\frac{d_{min}}{d_{max}}$ is explained by Eve's higher chances of capturing the transmission of the node lying at distance $d_{min}$.
3) The rates plotted in Figure 1 were found to be very close to those of a classifier that does not erase any received symbols, i.e. jammed symbols are always classified as belonging to a single node and forwarded to Eve's decoder.

### B. Experimental Results

We implemented our experiments on TinyOS [9] using TelosB motes [10], which have a built-in CC2420 radio module [11]. The CC2420 module uses the IEEE 802.15.4 standard in the 2.4GHZ band [12]. Our setup consists of four nodes, equivalent to Alice, Bob, Eve, and a Gateway module. Alice and Bob represent trusted nodes, while Eve represents the untrusted node that runs an energy-based classifier. The Gateway acts as a link between the sensor network and a PC running a java program. Our experiment is divided into cycles. During each cycle, the PC works as an orchestrator- *through the Gateway*- that determines, using a special message (*TRIGGER-MSG*), whether Alice should send alone, Bob sends alone, or both send concurrently. It also determines the power level used for transmission. These decisions are based on a user input transmission probability $P_t$. Upon receiving the broadcast TRIGGER-MSG, each trusted node will transmit a *DATA-MSG* while Eve will start to continuously read the value in the Received Signal Strength Indicator (RSSI) register (the RSSI value read by the CC2420 module is a moving average of the last 8 received symbols [11].). Eve will then transfer the RSSI readings from the memory buffer to the Gateway node which will forward them to the PC in an *RSSI-MSG*. For each cycle, the java program stores the received RSSI readings for further processing by the energy classifier (*implemented in MATLAB*). When transmitting data messages (*DATA-MSG*) from Alice or Bob, each node constructs a random payload of 100 bytes Using the RandomMlcg component of TinyOS, which uses the Park-Miller Minimum Standard Generator. Each symbol is *O-QPSK* modulated [12] representing 4 bits of the data. We also had to remove the CSMA-CA mechanism from the CC2420 driver in order to allow both Alice and Bob to transmit concurrently. Finally, it is worth noting that the orchestrator was used to overcome the synchronization challenge in our experimental set-up. In practical implementations, Bob (or Alice) could start jamming the channel upon receiving the Start of Frame Delimiter (SFD).

In our implementation of the energy classifier, the discrete nature of the transmit power levels was taken into consideration. First, the eavesdropper was given the advantage of having the classifier trained on a set of readings taken by running the experiment in the same environment and at the same node locations as those for which the classifier would be later used. In the training phase, our classifier is given prior information on the configuration, power levels selected for each node, and the measured RSSI readings at each cycle. It then finds the mean and variance of the measured RSSI values for each transmitted power level for Alice and Bob when each of them sends alone in a cycle. Any received symbol is classified as being transmitted by either of the communicating nodes. This choice is based on our third observation on the rates plotted in Figure 1. When running the classifier, it uses a *maximum likelihood* rule to determine the status of each transmitter in each cycle, i.e., the following expression is evaluated:

$$\frac{\max_i f_{A_i}(y)}{\max_i f_{B_i}(y)} \underset{B}{\overset{A}{\gtrless}} 1$$

and the symbol is classified accordingly, where $f_{X_i}(y)$ is the value of the approximated Gaussian distribution of measured RSSI values when source $X$ is the only transmitter and with power level $i$. In a practical implementation, the length of a cycle is the duration of a single symbol, hence, in our setup, the classifier bases its decision on a single RSSI reading. When evaluating our classifier, we use the transmission scenario indicating the actual status of the transmitters in each cycle and compare them with the classification results to obtain the probability of each possible misclassification event. We also assume that, in case of concurrent transmission, Eve can correctly decode the symbol received with higher signal power, as suggested in [13]. This assumption is used to calculate the values of $P_{e|(A,B)\to(A,B^c)}$ and $P_{e|(A,B)\to(A^c,B)}$. We also use the same set of data to train and run a classifier for the one-way TDM protocol described in section III-A. Here, we only consider cycles when Alice's transmitter was active and consider Bob's concurrent transmission as *jamming*. Finally, we evaluate it by finding $P_m$ and $P_{e|m}$.

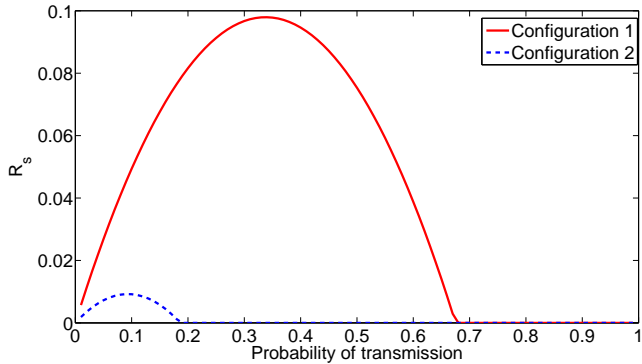

Fig. 2. $P_t$ vs. $R_s$ in different configurations for the randomized scheduling communication scheme, $R_s = [R_M\text{-}\max(R_{EA},R_{EB})]^+$.

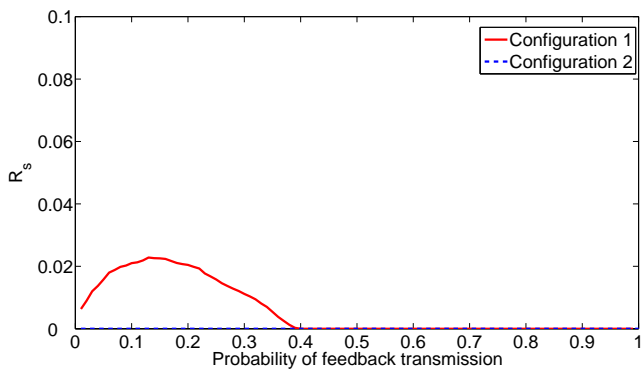

Fig. 3. $\beta$ vs. $R_s$ in different configurations for the one-way TDM scheme, $R_s = 0.5[R_M - R_E]^+$, we consider the case when Alice is the transmitter and Bob is the legitimate receiver.

We tested our scheme in a hallway environment, where only few scatterers exist (only the wall structure). We train, run, and evaluate our energy classifier, then use the resulting probabilities in the maximum rate expressions of section III to find the maximum achievable secrecy rate for different probability of transmission $P_t$ in case of the randomized scheduling communication scheme and probability of feedback transmission $\beta$ for the TDM scheme. Figures 2 and 3 show the maximum secrecy rate achieved for each of our two schemes. In Configuration 1, we set one of the Alice and Bob telosb nodes on top of each other, and set $d_{AE} = d_{BE} = 20ft$. In Configuration 2, we set $d_{AE} = 1ft$ and $d_{BE} = 20ft$. We note that the measured difference of received signal power values from both transmitting nodes was found to be 2dB and 19dB for Configurations 1 and 2 respectively. This implies that the maximum rates in Figures 2 and 3 should be compared to the value of $R_s$ in Figure 1 at $\frac{d_{min}}{d_{max}} = 0.79$ and 0.11 respectively. We believe that this difference between the theoretical and experimental results can be attributed to hardware differences and the deviation of the actual channel from the simplistic free space model used in our derivations. More specifically, we observe that the maximum secrecy rates for the randomized scheduling scheme in our experimental results is slightly lower than those calculated in Section IV-A at the mentioned distance ratios. The reason is Eve's enhanced ability to distinguish between the two sources of transmission due to the discrete nature of the selected transmit power values. Nevertheless, the experimental results establish the ability of our randomized scheduling and power allocation scheme for achieving perfect secrecy in practical scenarios akin to Body Area Networks where the distance between Eve and legitimate nodes will be larger than the inter-node distance **even if Eve is equipped with a very large receive antenna**.

V. CONCLUSIONS

This paper developed a novel physical layer approach for securing communication over two-way Gaussian channels in the presence of an eavesdropper with a very capable receive antenna. The underlying idea is to create an eavesdropper-ambiguity about the source of each symbol by randomizing the transmission schedule and power level. Our theoretical analysis revealed the ability of the proposed *randomization* approach to achieve relatively high **secure** transmission rates under mild conditions on the eavesdropper location. Our theoretical claims were further validated by extensive experimental results using IEEE 802.15.4-enabled sensor boards. Finally, we identified secure wireless body area networking (SW-BAN) as a natural candidate application for the proposed randomized scheduling and power allocation approach.

REFERENCES

[1] Yingbin Liang, H. V. Poor, and S. Shamai "Secure communication over fading channels," *IEEE Trans. Inf. Theory*, vol. 54, pp. 2470–2492
[2] Yingbin Liang and H. V. Poor "Secrecy capacity region of binary and Gaussian multiple-access channels,"
[3] T. Liu and S. Shamai "A note on secrecy capacity of the multi-antenna wiretap channel," *Submitted to IEEE Trans. Inf. Theory*
[4] C. E. Shannon, "Communication theory of secrecy systems," *Bell Syst. Tech. J.*, vol. 28, pp. 656–715, Oct. 1949.
[5] L. Lai, H. El Gamal, and H. V. Poor, "The wiretap channel with feedback: Encryption over the channel," *IEEE Trans. Inf. Theory*, vol. 54, no. 11, pp. 5059–5067, 2008.
[6] K. Srinivasan and P. Levis, "RSSI is under appreciated," *Third Workshop on Embedded Networked Sensors (EmNets)*, 2006.
[7] A. Wyner, "The wire-tap channel," *Bell Syst. Tech. J.*, vol. 54, pp. 1355–1387, 1974.
[8] I. Csiszár and J.Körner, "Broadcast channels with confidential messages," *IEEE Trans. Inf. Theory*, vol. 24, pp. 339–348, May 1978.
[9] J. Hill, R. Szewczyk, A. Woo, S. Hollar, D. Culler, and K. Pister, "System architecture directions for networked sensors," *Architectural Support for Programming Languages and Operating Systems*,vol. 35, pp. 93–104.
[10] "Telos data sheet."
[11] "Chipcon CC2420 datasheet".
[12] "IEEE 802.15.4 wireless medium access control (MAC) and physical layer (PHY) specifications for low-rate wireless personal area networks (LR-WPANs)".
[13] K. Whitehouse, A. Woo, F. Jiang, J. Polastre, and D. Culler, "Exploiting the capture effect for collision detection and recovery," *Proc. The Second IEEE Workshop on Embedded Networked Sensors, EmNetS-II,*, pp. 45–52, May 2005.s